\documentclass[preprint]{aastex}
\usepackage{emulateapj5,times,psfig}

\begin{document}
\title{Discovery of a kHz QPO in 2S~0918--549}

\author{Peter G. Jonker\altaffilmark{1}, Michiel van der
Klis\altaffilmark{1}, Jeroen Homan\altaffilmark{1}, Mariano
M\'endez\altaffilmark{1,2}, Jan van Paradijs\altaffilmark{1}, Tomaso
Belloni\altaffilmark{3}, Chryssa Kouveliotou\altaffilmark{4,5}, Walter
Lewin\altaffilmark{6}, Eric C. Ford\altaffilmark{1}}

\altaffiltext{1}{Astronomical Institute ``Anton Pannekoek'',
University of Amsterdam, and Center for High-Energy Astrophysics,
Kruislaan 403, 1098 SJ Amsterdam; peterj@astro.uva.nl,
michiel@astro.uva.nl, mmendez@fcaglp.fcaglp.unlp.edu.ar} 
\altaffiltext{2}{Facultad de Ciencias Astron\'omicas y
Geof\'{\i}sicas, Universidad Nacional de La Plata, Paseo del Bosque
S/N, 1900 La Plata, Argentina}
\altaffiltext{3}{Osservatorio Astronomico di Brera, via Bianchi 46,
I-23807 Merate, Italy; belloni@merate.mi.astro.it}
\altaffiltext{4}{Universities Space Research Association}
\altaffiltext{5}{NASA Marshall Space Flight Center, ES--84,
Huntsville, AL 35899}
\altaffiltext{6}{Department of Physics and Center for Space Research,
Massachusetts Institute of Technology, Cambridge, MA 02138}

\begin{abstract}
\noindent
We report the discovery of a kilohertz quasi--periodic oscillation
(kHz QPO) in the low--mass X--ray binary (LMXB) 2S~0918--549. The kHz
QPO has a frequency of 1156$\pm$9 Hz, a FWHM of 70$\pm$20 Hz, and a
fractional rms amplitude of 18\%$\pm$2\%. We also detected for the
first time a type I X--ray burst in this source. The compact object
must therefore be a neutron star. Finally, we were able to classify
the source as an atoll source exhibiting the full complement of
spectral/timing states of this class (island and banana). From the
peak burst flux an upper limit on the persistent luminosity can be
derived of 0.5\% of the Eddington luminosity, making 2S~0918--549 one of
the least luminous LMXBs showing a kHz QPO. We compare the fractional
rms amplitudes of the upper kHz QPO across the ensemble of LMXBs. We
find a strong anticorrelation with luminosity. In LMXBs with
luminosities $\sim$100 times lower than those of Z--sources, the
fractional rms amplitude is a factor $\sim$10 larger.

\end{abstract}


\section{Introduction}
\label{intro}
\noindent
Observations with the {\it Rossi X--ray Timing Explorer} (RXTE)
satellite have revealed the presence of several quasi--periodic
phenomena in the Fourier power spectra of low--mass X--ray binaries
(LMXBs). They occur on timescales similar to the dynamical timescale
near a neutron star, i.e. at frequencies around 1000 Hz.  Kilohertz
quasi--periodic oscillations (kHz QPOs) have been discovered in more
than 20 sources (van der Klis et al. 1996a,b; Strohmayer, Zhang, \&
Swank 1996; Strohmayer et al. 1996; see for the most recent review van
der Klis 2000). In several LMXBs nearly coherent oscillations were
discovered during type I X-ray bursts (Strohmayer et al. 1996; for a
review see Swank 2000); these burst oscillations presumably occur at
frequencies close to the neutron star spin frequency (Strohmayer et
al. 1996). Recently, another high frequency quasi-periodic phenomenon
was discovered in the power spectra of three LMXBs (Jonker, M\'endez,
\& van der Klis 2000). \par
\noindent
The kHz QPOs are nearly always found in pairs, although in a few
sources so far only one kHz QPO has been found (Zhang et al. 1998;
Marshall \& Markwardt 1999; Homan \& van der Klis 2000).  Their
frequencies vary over several hundred Hz on timescales of hours to
days; in the best studied sources, it is found that the frequency
separation of the twin kHz QPOs, $\nu_2-\nu_1$, decreases
monotonically by 50--100 Hz over the observed frequency range as the
frequency of the kHz QPO increases (see e.g. van der Klis et al. 1997;
M\'endez et al. 1998; M\'endez, van der Klis, \& van Paradijs 1998;
M\'endez \& van der Klis 1999); observations of the kHz QPO
frequencies in the other sources are consistent with this trend
(Jonker et al. 1998, Psaltis et al. 1998). \par
\noindent
Various models exist for these QPOs. Immediately after their discovery
a beat frequency model was proposed (Strohmayer et al. 1996), of which
the sonic--point model is the most current (Miller, Lamb, \& Psaltis
1998; revised by Lamb \& Miller 2000). Later, Stella \& Vietri (1999)
proposed the relativistic precession model (but see Markovi\'c \& Lamb
2000) and Osherovich \& Titarchuk (1999) introduced the
two--oscillator model. In the latter model the QPO at $\nu_1$ (the
lower kHz QPO) occurs at the Keplerian frequency of material orbiting
the neutron star, whereas in the other two models the QPO at $\nu_2$
(the upper kHz QPO) is the one that is Keplerian. Recently, Psaltis \&
Norman (2000) proposed the transition layer model in which the twin
kHz QPO peaks arise due to disk oscillations at frequencies close to
those predicted by the relativistic precession model.  \par
\noindent 
In this Letter, we report the discovery of a kHz QPO and a type I
X--ray burst in 2S~0918--549. The source turns out to be one of the
least luminous kHz QPO sources, however, the fractional rms amplitude
of the QPO is quite high.  We study the dependence of the fractional
rms amplitudes of the upper kHz QPO on source luminosity across the
ensemble of LMXBs and show that they are strongly anticorrelated.

\section{Observations and analysis}
\label{analysis}
\noindent
We have used observations obtained with the proportional counter array
(PCA; Jahoda et al. 1996) onboard the {\it RXTE} satellite (Bradt,
Rothschild, \& Swank 1993). A log of the observations and the average
source count rates at the time of the observation can be found in
Table~\ref{obs_log}. In total $\sim$75 ksec of data were used in our
analysis. Data were obtained in four different modes. The Standard 1
and 2 modes, which are always operational, respectively provide data
with a time resolution of 0.125 seconds in just one energy bin and 16
seconds in 129 energy bins covering the effective PCA 2 to 60 keV
range. Additionally, data were obtained using the mode providing the
highest time resolution ($\sim 1 \mu s$), combined with the full
energy resolution {\it RXTE} can provide (256 energy bins covering the
2 to 60 keV; the GoodXenon mode). This mode saturates on count rates
higher than $\sim$8000 counts per second.\par
\noindent
In order to follow spectral variations we created color--color
diagrams (CDs) by plotting a soft color vs. a hard color. The soft
color was defined as the ratio between the count rates in the 3.5--6.4
keV and 2--3.5 keV energy band; the hard color was defined as that
between the 9.7--16.0 keV and 6.4--9.7 keV energy band. The data were
background subtracted but no dead--time corrections were applied (the
dead--time is less than $\sim$2\%). The colors of the 1997
observations were corrected for PCA gain changes, using Crab
observations obtained in May 1997, and May 2000. Assuming the Crab is
constant, observed color changes of the Crab between the observations
of May 1997 and May 2000 should reflect changes of the instrument's
spectral response. The observed changes in Crab were $\sim$6\% in the
soft color and $\sim$1\% in the hard color. The colors were converted
to those which would have been detected by the PCA on May 2000 (cf. di
Salvo et al. 2000). Residual color errors due to the differences
between the spectra of Crab and 2S~0918--549 are less than the
statistical errors.  \par
\noindent
We calculated Fast Fourier Transforms with a Nyquist frequency of 4096
Hz of data segments of 16 seconds (2--60 keV). The power density
spectra were averaged and fitted with a function consisting of
Lorentzians (to represent peaks arising in the power spectrum due to
QPOs), a constant to account for the power due to Poisson counting
statistics, and an exponentially cutoff power law to represent the
noise component apparent at low frequencies.  Errors on the fit
parameters were calculated using $\Delta\chi^2=1.0$ (1$\sigma$ single
parameter). The 95\% confidence upper limits were determined using
$\Delta\chi^2=2.71$.

\section{Results}
\label{result}
\begin{figure*}
\centerline{\psfig{file=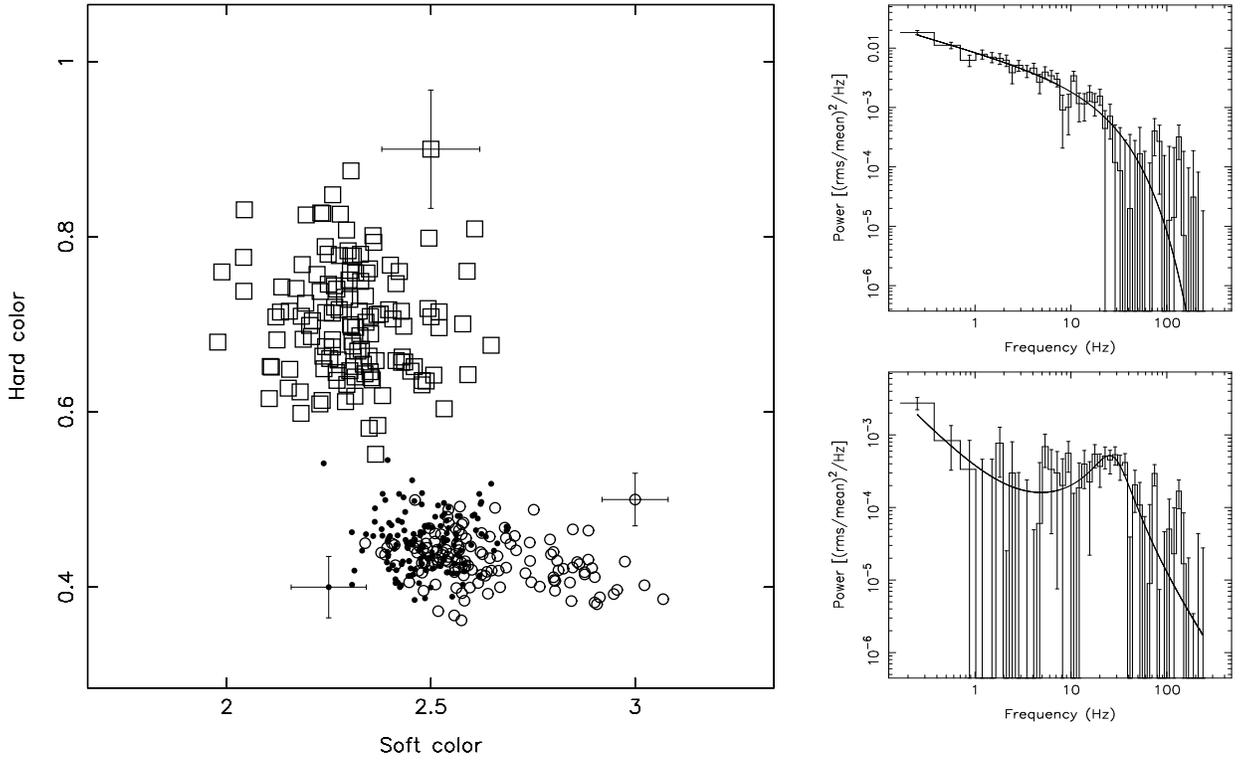,width=20cm,angle=90}}\figcaption{{\it
Left:} Color--color diagram of 2S~0918--549. The data were background
subtracted, the colors of the 1997 observations were converted to the
May 2000 observations using the Crab as a reference (see text), and
the burst was removed. No dead--time corrections have been applied
(the dead--time is $\sim$2\%). The squares and circles are 256~s and
128~s averages, respectively. Typical error bars are shown for each
category. The pattern resembles the Island state (squares) and the
banana branch (open and filled circles) of atoll sources (see Hasinger
\& van der Klis 1989). The data in which the kHz QPO was present are
represented by filled circles. {\it Right, top:} Rms normalized,
Poisson noise subtracted average power spectrum of the observations
1--10 (squares). Strong band limited noise is present, establishing
the identification of the Island state. The solid line is the best fit
to the data with a cut--off powerlaw. {\it Right, bottom:} Rms
normalized, Poisson noise subtracted average power spectrum of the
observations 11--13 (circles). Banana state power law noise and a
peaked noise component are present. The solid line is the best fit
to the data with a powerlaw and a Lorentzian component. \label{cd}}
\end{figure*}

\noindent
The color--color diagram of 2S~0918--549 (see Fig.~\ref{cd};
left) resembles that of an atoll source. The squares represent the
average of 256~s of data from observations 1--10, the filled circles
are 128~s averages from observation 11, and the open circles are 128~s
averages from observations 12 and 13. Together with the timing
properties (described below) it is clear that the squares represent
the Island state and the open and filled circles the banana branch of
an atoll source.

\noindent
The average power spectrum of observations 1--10 combined (see
Fig.~\ref{cd}, right, top) shows a strong band limited noise
component, which could be described using an exponentially cutoff
power law with a fractional rms amplitude of 26\%$\pm$1\% (integrated
over 0.1--100 Hz, 2--60 keV), a power law index of 0.47$\pm$0.05, and
a cutoff frequeny of 21$\pm$6 Hz. We derived a 95\% confidence upper
limit on the presence of a kHz QPO with a full-width at half maximum
(FWHM) 100 Hz in the range of 300--1200 Hz of $\sim$30\% (rms).  \par
\begin{figure*}
\centerline{
\psfig{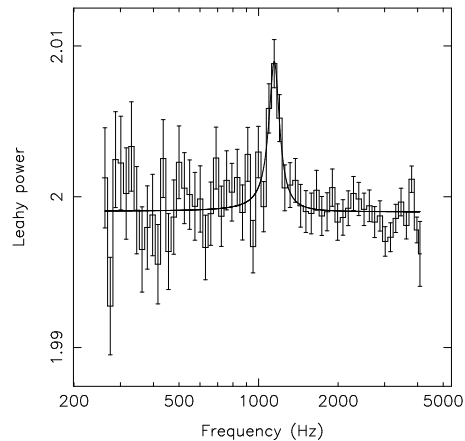}}
\figcaption{Average 4.0--18.0 keV power density spectrum of observation 11,
normalized after Leahy et al. (1983), showing the kHz QPO at
1156$\pm$9 Hz ($6.1\sigma$ single trial). The solid line represents
the best fit to the data. \label{kHz}}
\end{figure*}

\noindent
We discovered a kHz QPO at 1156$\pm$9 Hz, with a FWHM of 68$\pm$23 Hz,
and a fractional rms amplitude of 18\%$\pm$2\% (2--60 keV) in the
average power spectrum of observation 11 (see Fig.~\ref{kHz}).  In
addition, the average power spectrum of observation 11 (Fig.~\ref{cd},
right, bottom) shows a peaked noise component at a frequency of
25$\pm$2 Hz, with a fractional rms amplitude of 17\%$\pm$2\%, and a
FWHM of 22$\pm$6 Hz. To investigate the peak noise component and the
kHz QPO further, we also calculated power spectra using data in the
4.0--18.0 keV energy band. The values measured in the 4.0--18.0 keV
energy band are the same within the errors, although the significance
of the detection of the kHz QPO is somewhat higher (6.1$\sigma$
compared to 5.7 $\sigma$ in the 2--60 keV band). We subdivided the
4.0--18.0 keV power spectra of observation 11 in two parts. The
frequency of the kHz QPO increased from 1126$\pm$9 Hz in the first
part of observation 11 to 1218$\pm20$ Hz at the end. The source moved
further up the banana branch, from an average hard and soft color of
0.452$\pm$0.003 and 2.49$\pm$0.01 to 0.431$\pm$0.003 and
2.64$\pm$0.01, respectively. The background subtracted count rate in
the 4.0--18.0 keV energy band, the FWHM, and the fractional rms
amplitude are consistent with being the same in the two parts. The
properties of the peaked noise component were consistent with being
the same in the two selections. \par
\noindent
\noindent
No power spectral components were detected in the average power
spectrum of observations 12 and 13 combined. We derived an upper limit
on the presence of a 100 Hz wide kHz QPO of $\sim$13\% rms in the
range between 300--1200 Hz in both the 2--60 keV and the 4.0--18.0 keV
energy band, and an upper limit of 5\%--6\% on the presence of a power
law component at frequencies in the range of 0.1 to 100 Hz in both the
2--60 keV and the 4.0--18.0 keV energy band, with a power law index of
1.0. \par
\begin{figure*}
\figurenum{3} \centerline{\psfig{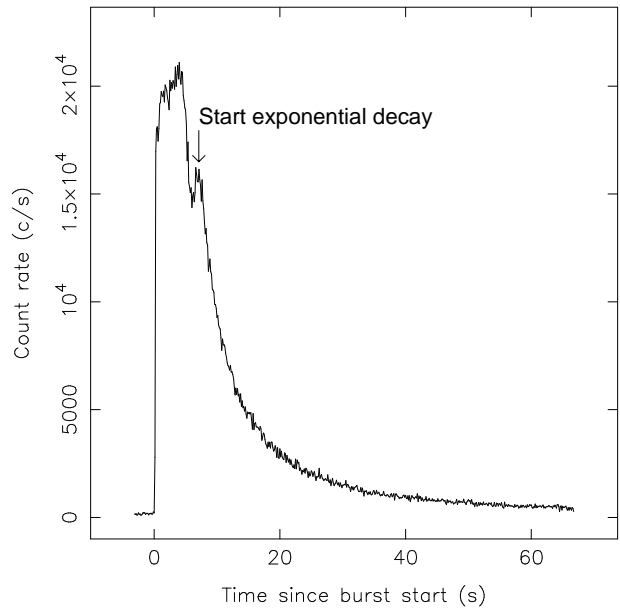}}
\figcaption{Lightcurve showing the profile of the type I X-ray burst
occurring during observation 11. The X-axis denotes the time since the
start of the burst. Note the complex burst profile and the large
increase in count rate during the burst compared with the persistent
emission. The count rate is that of the three detectors that were
operational only.}
\label{lc_burst}
\end{figure*}

\noindent
During observation 11 a type I X--ray burst occurred (see
Fig.~\ref{lc_burst}). This is the first detection of a burst in
2S~0918--549, and establishes the compact object as a neutron star. As
apparent in Fig.~\ref{lc_burst} the burst profile was complex, with a
smooth exponential decay with an e-folding time of 8.95$\pm$0.05~s
beginning $\sim10$ seconds after the initial rise. Due to the very
high count rates of more than 20000 counts per second (3 detector
count rate) the GoodXenon mode saturated. The total count rate
displayed in Fig.~\ref{lc_burst} is for the 0.125 s Standard 1 data
which were not affected by this saturation. During the part of the
decay where the count rate was less than 8000 counts per second
spectral analysis shows cooling of the blackbody spectral
component. Using the 16~s Standard 2 mode we derive a flux for the
16~s bin containing the burst peak of $3.5 \times 10^{-8}$ erg
cm$^{-2}$ s$^{-1}$ (2-20 keV). Since the mean count rate in this 16~s
time bin is a factor of $\ga$ 2.5 lower than the peak count rate as
measured using the 1/8~s bin of the Standard 1 observations, we use
this factor to improve our estimation of the peak burst flux. We did
not correct for the deadtime effects which become important at the
count rates at the peak of the burst ($\sim$25\%), this provides a
lower limit on the peak flux. The pre--burst flux is less than $4.2
\times 10^{-10}$ erg cm$^{-2}$ s$^{-1}$ (2-20 keV). So, the persistent
emission of the source when it is on the lower banana branch is less
than 0.5 percent of the Eddington luminosity. Assuming the peak burst
flux to be isotropic and at or below the Eddington limit of $2.5
\times 10^{38}$ erg s$^{-1}$ we derive an upper limit to the distance
of 4.9 kpc.

\section{Discussion}
\noindent
We discovered a kHz QPO at a frequency of $\sim$1156 Hz with a
fractional rms amplitude of nearly 20\% (2--60 keV) in the X--ray
emission of the LMXB 2S~0918--549. For the first time in this source a
type I X--ray burst was detected, establishing the nature of the
compact object in 2S~0918--549 as a neutron star. We derived an upper
limit to the persistent luminosity when the source is on the lower
banana branch of 0.005 $L_{Edd}$ making 2S~0918--549 one of the least
luminous LMXBs in which a kHz QPO has been observed. The spectral and
timing properties indicate that this source is a low--luminous atoll
source, displaying the full complement of spectral and timing states
(island state and a full banana branch). Similar to other kHz QPO
sources, e.g. M\'endez et al. (1999), the kHz QPO was only observed
when the source was on the lower part of the banana branch. Its
frequency changed from $1126\pm$8 Hz to 1218$\pm$20 Hz, when the
source moved further on the banana branch, similar to other atoll
sources (e.g. M\'endez et al. 1999). Its frequency is similar to the
frequency of the {\em upper} kHz QPO in other atoll sources in a
similar lower--banana state (e.g. di Salvo et al. 2000, M\'endez et
al. 1999). The frequency of the lower kHz QPO is usually lower in that
part of the CD.  Furthermore, when both QPOs of the twin kHz QPOs are
detected simultaneously the frequency of the lower kHz QPO has never
been found at frequencies larger than $\sim$900 Hz in any LMXB (see
van der Klis 2000). Therefore, we conclude that this kHz QPO is most
likely the upper peak of the kHz QPO pair which is detected in most
other LMXBs. \par
\noindent
After 4U~0614+109, 2S~0918--549 is another example of a very low
luminous LMXB whose kHz QPOs are strong, whereas in the most luminous
LMBXs, the Z--sources, the kHz QPOs tend to be weak (see also the
review of van der Klis 2000). To further investigate this, we compared
the fractional rms amplitudes of the kHz QPO in 2S~0918--549 with
those of the upper kHz QPO in other LMXBs. In order to account for the
fact that in a given source the fractional rms amplitude of the kHz
QPO changes as its frequency changes we measured the fractional rms
amplitude at both the lowest and highest frequency at which the upper
kHz QPO is found. From extensive studies for several sources it is
clear that the fractional rms amplitude of the upper kHz QPO decreases
monotonically as the frequency of the upper peak increases
(e.g. Wijnands et al. 1997, Jonker et al. 1998, Wijnands et al. 1998,
M\'endez, van der Klis, \& Ford 2000). This ensures that the method we used
will yield a good indication of the range in fractional rms amplitude
spanned in each source. The sample we studied here contains only those
sources for which a significant detection of the upper kHz QPO in the
5--60 keV energy range was possible (4U~0614+09; Ford et al. 1996;
4U~1915--05, Boirin et al. 2000; 4U~1728--34, Strohmayer, Zhang, \&
Swank 1996; 4U~1608--52, M\'endez et al. 1998; 4U~1702--42, Markwardt
et al. 1999; 4U~1636--53, Zhang et al. 1996; 4U~1820--30, Smale et
al. 1996; XTE~J2123--058, Homan et al. 1999; KS~1731--260, Wijnands \&
van der Klis 1997), or in case of the Z-sources, the 5-60 keV
fractional rms amplitude that was reported in the literature
(Cyg~X--2, Wijnands et al. 1998; GX~17+2, Wijnands et al. 1997;
GX~340+0, Jonker et al. 1998). Sco~X--1 was not included since the
fractional rms amplitude of the QPOs can not be calculated accurately
at present due to the severe dead-time effects present in this
source. We calculated the 5--60 keV fractional rms amplitude in order
to avoid as much as possible systematic effects on the derived
fractional rms amplitudes of the peaks due to the different $N_H$
values towards the various sources.\par
\begin{figure*}
\figurenum{4}
\centerline{\psfig{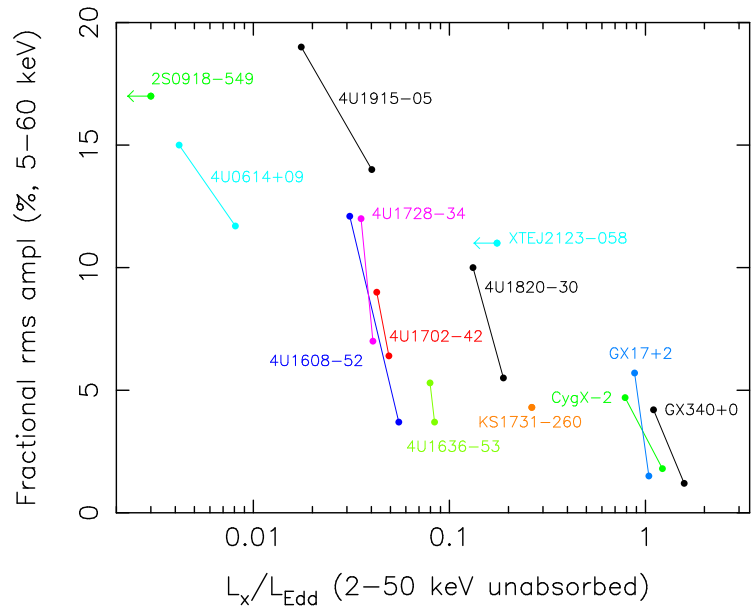}}
\figcaption{Luminosity versus fractional rms amplitude (5--60 keV) of
the upper kHz QPO in various LMXBs. Apparent is the decrease in
fractional rms amplitude as the source luminosity increases, both
within one source and between sources.}
\label{rmsen}
\end{figure*}

\noindent
The range in fractional rms amplitude of the upper kHz QPO in each
source is plotted in Fig.~\ref{rmsen} as a function of the
simultaneously measured luminosity (for the distances and the method
used to calculate these luminosities see Ford et al. 2000 and
references therein; the distance to 4U~1915--05 was taken from Smale
et al. 1988). It is apparent that the upper kHz QPOs detected in
low--luminosity sources have relatively large fractional rms
amplitudes, when compared with the fractional rms amplitude of the kHz
QPO in the bright LMXBs (e.g. the Z-sources). The frequencies of the
end--points for each source are given in Table~\ref{freqs}. \par
\noindent
Van Paradijs \& van der Klis (1994) found that the hardness, defined
as the ratio of counts between the 40--80 keV band and the 13--25 keV
band, is in neutron star LMXBs correlated with the X--ray
luminosity. This suggests that the X--ray flux variability of the
source at the kHz QPO frequency is correlated with the hardness in
that band (see also Ford et al. 1997). Within a source a similar
correlation between the fractional rms amplitude of the upper kHz QPO
and the hardness of the energy spectrum has been found; the harder the
source the higher the fractional rms amplitude of the upper kHz QPO
(e.g. 4U~0614+091, Ford et al. 1997, van Straaten et al. 2000;
4U~1608--52, M\'endez et al. 1999; Aql~X--1, Reig et al. 2000;
4U~1728--34, Di Salvo et al. 2000; Cyg~X--2, Wijnands et al. 1998;
GX~17+2, Wijnands et al. 1997; GX~340+0, Jonker et al. 1998).\par
\noindent
A two--component accretion flow (radially and through a disk, Gosh and
Lamb 1979) has been considered to explain that the frequency of the
kHz QPO does not depend on source luminosity (Ford et al. 2000).  The
observed anticorrelation between the fractional rms amplitude of the
upper kHz QPO and the source luminosity cannot solely be explained by
an increase of an unmodulated part of the accretion flow towards
higher source luminosity, since the increase in luminosity is too
large compared with the decrease in fractional rms
amplitude. Therefore, this cannot explain the findings of Ford et
al. (2000). The same conclusion was drawn by M\'endez et al. (2000) on
other grounds. It is important to consider the anticorrelation
presented in this work and the absence of a correlation reported by
Ford et al. (2000) in mechanisms producing kHz QPOs.

\acknowledgments This work was supported in part by the Netherlands
Organization for Scientific Research (NWO) grant 614-51-002. This
research has made use of data obtained through the High Energy
Astrophysics Science Archive Research Center Online Service, provided
by the NASA/Goddard Space Flight Center. This work was supported by
NWO Spinoza grant 08-0 to E.P.J.van den Heuvel. WHGL gratefully
acknowledges support by NASA.

\begin{deluxetable}{ccccc}
\tablecaption{Log of the observations of 2S~0918--549 used in this
analysis. The count rates in column 5 are the average, background
subtracted, 2--60 keV count rates obtained as if 5 PCUs were always
operational. \label{obs_log}}

\tabletypesize{\normalsize}
\tablecolumns{5}
\tablewidth{0pc}
\tablehead{
\colhead{} & \colhead{Observation} & \colhead{Date \&} &
\colhead{Amount of data} & \colhead{Average count rate}\\
\colhead{} & \colhead{ID} & \colhead{Start time
(UTC)} & \colhead{(ksec)} & \colhead{cnts/sec (2--60 keV)}}
\startdata               
1 & 20071-11-01-02  &	02-05-1997	23:57 & $\sim$4.2 & $\sim$36\\
2 & 20071-11-01-03  &	03-05-1997	02:15 & $\sim$2.1 & $\sim$36\\
3 & 20071-11-01-04  &	03-05-1997	06:42 & $\sim$1.8 & $\sim$45\\
4 & 20071-11-01-05  &	03-05-1997	08:26 & $\sim$1.4 & $\sim$50\\
5 & 20071-11-01-06  &	04-05-1997	08:27 & $\sim$1.5 & $\sim$53\\
6 & 20071-11-01-00  & 	04-05-1997	22:41 & $\sim$11.3 & $\sim$47\\
7 & 20071-11-01-01  &	05-05-1997	22:16 & $\sim$6.0 & $\sim$48\\
8 & 20071-11-01-07  &	09-05-1997	10:10 & $\sim$2.4 & $\sim$54\\
9 & 20064-06-01-00  &	15-08-1997	01:38 & $\sim$4.8 & $\sim$84\\
10 & 20064-06-02-00 &	21-09-1997	22:51 & $\sim$5.0 & $\sim$68\\
11 & 50060-01-01-00 &	12-05-2000	12:24 & $\sim$30.0 & $\sim$135\\
12 & 50060-01-01-01 &	14-05-2000	19:43 & $\sim$2.1 & $\sim$187\\
13 & 50060-01-01-02 &	15-05-2000	04:59 & $\sim$2.4 & $\sim$208
\enddata
\end{deluxetable}

\begin{deluxetable}{lll}
\tablecaption{The lowest and highest frequency of the upper kHz QPO
for each source used to create Fig. 4. \label{freqs}}

\tabletypesize{\normalsize}
\tablecolumns{5}
\tablewidth{0pc}
\tablehead{
\colhead{Source name} & \colhead{Lowest frequency} & \colhead{Highest
frequency} \\
\colhead{} & \colhead{upper kHz (Hz)} & \colhead{upper kHz (Hz)}}
\startdata               
2S~0918--549  & 1156$\pm$9 & 1156$\pm$9\\
4U~0614+09    & 421$\pm$13  & 1161$\pm$5\\
4U~1915--05   & 542$\pm$14  & 1013$\pm$8\\
4U~1608--52   & 883$\pm$4  & 1091$\pm$8\\
4U~1728--34   & 552$\pm$7  & 1139$\pm$13\\
4U~1702--42   & 1000$\pm$9  & 1059$\pm$11\\
4U~1636--53   & 1152$\pm$4  & 1213$\pm$11\\
4U~1820--30   & 675$\pm$4  & 1050$\pm$4\\
XTE~J2123--058& 1129$\pm$8  & 1129$\pm$8\\
KS~1731--260  & 1176$\pm$3  & 1176$\pm$3\\
Cyg~X--2      & 731$\pm$19  & 1007$\pm$15\\
GX~17+2       & 645$\pm$9  & 1086$\pm$8\\
GX~340+0      & 535$\pm$70  & 840$\pm$20
\enddata
\end{deluxetable}

\end{document}